\documentclass{jfm}
\usepackage{xcolor}
\usepackage{array}
\usepackage{multirow}
\usepackage{tabularx}

\begin{document}

\newtheorem{lemma}{Lemma}
\newtheorem{corollary}{Corollary}

\shorttitle{Transitions near the onset of low Prandtl-number rotating magnetoconvection} 
\shortauthor{Banerjee et al.} 

\title{Overstable rotating convection in presence of vertical magnetic field}

\author
 {
 Ankan Banerjee\aff{1},
 Manojit Ghosh\aff{1},
  Lekha Sharma \aff{1},
  \and
  Pinaki Pal\aff{1}
  \corresp{\email{pinaki.math@gmail.com}}
  }

\affiliation
{
\aff{1}
Department of Mathematics, National Institute of Technology, Durgapur~713209, India
}

\maketitle

\begin{abstract}
We present the results of our investigation on nonlinear overstable rotating magnetoconvection (RMC) in presence of vertical external magnetic field. We focus on the dynamics appearing near the onset of convection by varying the system control parameters, namely, the Taylor number ($\mathrm{Ta}$), the Chandrasekhar number ($\mathrm{Q}$) and the Prandtl number ($\mathrm{Pr}$) in the ranges $750\leq\mathrm{Ta}\leq10^6$, $0 < \mathrm{Q} \leq 10^3$ and $0 < \mathrm{Pr} \leq 0.5$. Three dimensional (3D) direct numerical simulations (DNS) of the governing equations and low-dimensional modeling of the system are performed for this purpose. Extensive DNS in the specified parameter space shows two qualitatively different onsets depending on $\mathrm{Ta}$, $\mathrm{Q}$ and $\mathrm{Pr}$. In the first one, bistability appears at the onset, where both subcritical and supercritical convection coexist, while only supercritical convection is observed in the second one. Analysis of the low-dimensional model reveals that a supercritical Hopf bifurcation is responsible for the supercritical onset and a subcritical pitchfork bifurcation is responsible for the subcritical onset. It is also observed that appearance of subcritical convection at the onset has strong dependence on all three control parameters $\mathrm{Ta}$, $\mathrm{Q}$ and $\mathrm{Pr}$. The scenario of subcritical convection is found to disappear as $\mathrm{Pr}$ is increased for fixed $\mathrm{Ta}$ and $\mathrm{Q}$. However, most striking findings of the investigation is that the increment in $\mathrm{Ta}$ for fixed $\mathrm{Q}$ and $\mathrm{Pr}$ opposes the subcritical convection, whereas the increment in $\mathrm{Q}$ for fixed $\mathrm{Ta}$ and $\mathrm{Pr}$ favors it.  This is in sharp contrast with the earlier results reported in RMC. 

\end{abstract}

\section{Introduction}\label{sec:introduction}
Rotating magnetoconvection (RMC) generally refers to the convective motion of a fluid in simultaneous presence of rotation and magnetic field. It is common to many geophysical and astrophysical applications and plays an important role in determining the dynamics of these systems~\citep{eltayeb:1975, Roberts:2000, Olson:2001, glatzmaier_book:2013}. For example, RMC is seen in the Earth's liquid outer core and outer $30\%$ of the solar radius~\citep{glatzmaier_book:2013} and in both the cases it strongly affects the transport of various quantities including mass, momentum and energy. Rotation about an axis introduces the Coriolis and centrifugal forces while motion of electrically conducting fluids in presence of the magnetic field generates the Lorentz force. Presence of these forces together with the arbitrariness of the system geometry makes RMC a more complex problem. Therefore, researchers rely on some simplified models to get insight into these phenomena.

The omnipresence of thermal convection in various natural as well as man-made systems has always inspired the researchers to look into the basic physics underneath it. The fluid flow between two parallel plates heated from below and cooled from above provides a simplified model of convection, classically known as the Rayleigh-B\'{e}nard convection (RBC). Since its introduction in early twentieth century, RBC has been studied over a century to get insight into various phenomena such as instabilities, patterns-dynamics, chaos, turbulence etc. and yet it is an active area of research~\citep{bodenschatz:ARFM_2000, ahlers:RMP_2009, mkv:book}. Not only these studies have enriched the knowledge of RBC but also contributed significantly in the development of many related subjects including hydrodynamic stability~\citep{chandra:book, drazin:book}, nonlinear dynamics~\citep{busse:1978, getling:book, manneville_book:2010} and pattern formation~\citep{hoyle_pattern_formation:book}.

The dynamics of RBC is controlled by two dimensionless parameters namely, the Rayleigh number ($\mathrm{Ra}$, vigor of the buoyancy) and Prandtl number ($\mathrm{Pr}$, ratio of kinematic viscosity and thermal diffusivity of the fluid). Note that, the Prandtl number is a characteristic of the fluids which measures the relative strength of two nonlinear terms present in the momentum and energy equations respectively. For low-$\mathrm{Pr}$ fluids, which are more relevant in the context of geophysical and astrophysical flows (e.g., for Earth's core $\mathrm{Pr} \sim 10^{-1}-10^{-2}$ whereas for stellar interiors $\mathrm{Pr} \sim 10^{-5}-10^{-9}$), velocity nonlinearity in the momentum equation dominates the nonlinearity present in the energy equation. As a result, vertical vorticity is generated very close to the convection threshold and rich bifurcation structures are observed there~\citep{busse:JFM_1972, meneguzzi:1987, clever:POF_1990, nandu:2016}. 

Low-$\mathrm{Pr}$ fluids are also electrically conducting and presence of magnetic field strongly affects their motion. RBC in presence of magnetic field, generally called the Rayleigh-B\'{e}nard magnetoconvection (RBM) serves as a simplified model to study the magnetoconvective instabilities in great detail~\citep{proctor:book}. Two more control parameters appear in the description of RBM, namely, the Chandrasekhar number ($\mathrm{Q}$, measures strength of the Lorentz force) and the magnetic Prandtl number ($\mathrm{Pm}$, ratio of kinematic viscosity and magnetic diffusivity of the fluid). It is well established that presence of magnetic field in horizontal directions does not affect the primary instability~\citep{busse:1983, fauve:1984}. However, it does influence the secondary instabilities significantly~\citep{fauve_prl:1984, nandu:2015}. On the other hand, presence of magnetic field in the vertical direction strongly affects the primary instability as well as the secondary instabilities~\citep{knobloch_JFM:1981, busse:1982, busse:1989a, arnab:2014}. 

Like magnetic field, rotation also has nontrivial effects on the convective motion of fluids. Rotation about the vertical axis exerts the Coriolis and centrifugal forces and both of them act along the horizontal plane. The centrifugal force generally brings in the large scale circulation in the flow and modifies the basic temperature profile~\citep{maity:2014}. On the other hand, the Coriolis force which is measured by the Taylor number ($\mathrm{Ta}$) linearly couples the vertical velocity and vertical vorticity which delays the onset, affects the heat flux and breaks the mirror symmetry of the system even at small rotation rates~\citep{maity:2014}. Moreover, presence of rotation also facilitates the subcritical motion of the fluids~\citep{veronis:1966, kaplan:2017}.

It is found that both rotation and magnetic field provide certain properties of elasticity to the fluid so that it can sustain appropriate modes of wave propagation~\citep{moore:1973}. This enables the overstable oscillatory motion of the fluids in which the convection sets in as time-dependent purely oscillatory motion. As a result, the `principle of exchange of stability' becomes invalid in this case. The overstable oscillatory convection has been studied in great detail either in presence of rotation~\citep{pharasi:2013, maity:2014} or magnetic field~\citep{proctor:book, mondal:2018} due to its relevance to various astrophysical applications such as sunspot, accretion discs and planetary interiors~\citep{roberts:1976, latter_MNRA:2015}. It is observed that for rotating convection, the overstable oscillatory convection occurs only when $\mathrm{Ta}$ crosses a critical value depending on $\mathrm{Pr}$~\citep{chandra:book}. In addition, the scenario of overstability vanishes for $\mathrm{Pr} \geq 0.667$ which eligible the low-$\mathrm{Pr}$ fluids to be the ideal candidate to study the overstable oscillatory convection. On the other hand, occurrence of overstability in magnetoconvection depends on both $\mathrm{Pr}$ and $\mathrm{Pm}$. It has been found that overstability occurs in magnetoconvection with vertical magnetic field only when $\mathrm{Pm} > \mathrm{Pr}$~\citep{proctor:book}. Interestingly, presence of rotation together with the magnetic field loosens such restriction and overstability is observed even when $\mathrm{Pm} < \mathrm{Pr}$.

However, overstability in RMC is much less studied despite of the fact that RMC mimics the geophysical and astrophysical situations more closely.~\cite{chandra:book} first performed the linear stability analysis of the problem by considering RBC in simultaneous presence of rotation about the vertical axis and an external uniform vertical magnetic field. Later,~\cite{nakagawa:1957,nakagawa:1959} carried out experiments on RMC and validated the theoretical findings of Chandrasekhar with the experimental observations. In a subsequent study,~\cite{eltayeb:1975} investigated the RMC with various orientations of rotation and magnetic field together with different boundary conditions by performing asymptotic analysis (both $\mathrm{Ta,Q} \rightarrow \infty$) to determine some well defined scaling laws for the onset of convection. Roberts and Jones studied the RMC further with horizontal magnetic field to determine the preferred mode of convection using linear theory in the limit $\mathrm{Pr} \rightarrow \infty$~\citep{Roberts:2000, Jones:2000}.~\cite{podvigina_FD:2008} also studied the RMC by performing linear stability analysis and identified the parameter space where convection is manifested as overstable oscillatory instability.~\cite{Eltayeb:2013} further studied RMC using linear theory in which both rotation and magnetic field are horizontal and inclined at an angle $\phi$ for better understanding of the role of viscosity, the electrical conductivity of the boundary and the interaction among all possible wave motions. Therefore, it is evident from the literature that most of the studies on RMC are based on linear theory and its nonlinear aspects have remained largely unexplored. Recently,~\cite{banerjee_PRE:2020} performed direct numerical simulation (DNS) and low-dimensional modeling of the system to explore the dynamics appearing near the onset of RMC with horizontal magnetic field by considering the full nonlinear problem. They assumed quasi-static approximation and reported simultaneous appearance of subcritical and supercritical convection at the onset. They also showed that appearance of subcritical convection at the onset is independent of $\mathrm{Q}$ and depends only on $\mathrm{Ta}$ and $\mathrm{Pr}$. However, the effect of vertical magnetic field on overstable oscillatory rotating convection has remained unanswered. In the present work, we address this issue by performing DNS of the governing equations and low-dimensional modeling of the full nonlinear problem. Our study reveals rich bifurcation structures near the onset depending on the control parameters $\mathrm{Ta}$, $\mathrm{Q}$ and $\mathrm{Pr}$. But, the effect of $\mathrm{Ta}$ and $\mathrm{Q}$ on these bifurcations are found to be quite surprising and counter-intuitive. We hope these findings will contribute significantly to the future investigations on RMC. 

The rest of the paper is designed as follows. The mathematical description of the problem is presented in Section~\ref{sec:system}. Section~\ref{sec:lsa_nm} discusses about the linear stability analysis and numerical techniques we have used in this work. Section~\ref{sec:results} is devoted to the results and discussions. Finally, some concluding remarks are drawn in Section~\ref{sec:conclusion} with some future perspectives. 

\section{Problem configuration}\label{sec:system}
\begin{figure*}
\includegraphics[height=!, width =\textwidth]{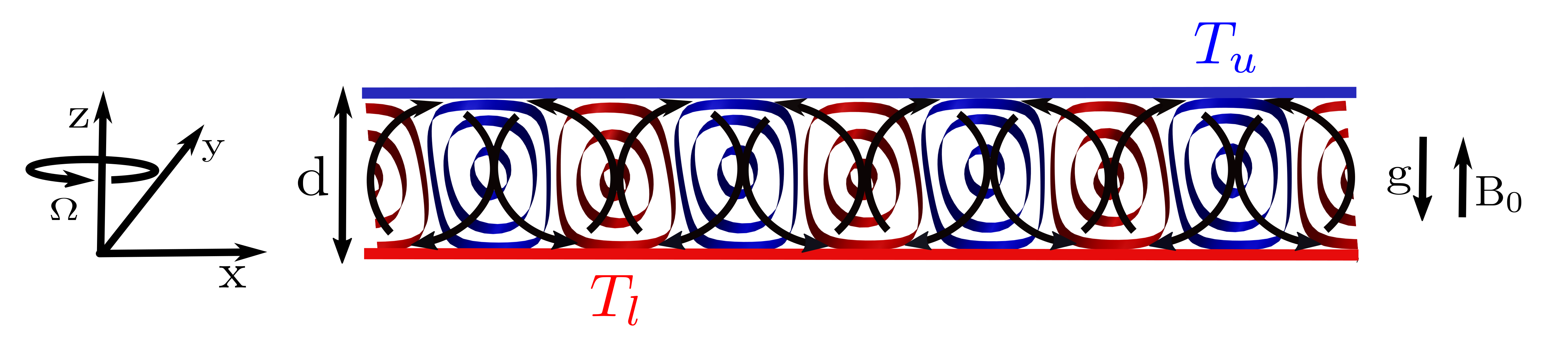}
\caption{Schematic representation of the rotating magnetoconvection (RMC) set-up.}
\label{rbc:setup}
\end{figure*}

Plane layer Rayleigh-B\'{e}nard convection (RBC) set-up has been considered for the study. An infinitely extended thin horizontal layer of electrically conducting Boussinesq fluid of thickness $d$, kinematic viscosity $\nu$, coefficient of volume expansion $\alpha$, magnetic diffusivity $\lambda$, and thermal diffusivity $\kappa$ is confined between two thermally conducting parallel plates. The lower plate is heated uniformly and the upper plate is cooled to maintain a vertical temperature gradient \( \beta = (T_l - T_u) /d \) across the fluid layer, $T_l$ and $T_u$ being the temperatures of the lower and upper plates respectively. The fluid is subjected to an external uniform vertical magnetic field $\mathbf{B_0}\equiv (0,0,B_0)$. The whole system is rotating uniformly about the vertical axis with angular velocity $\Omega$. A schematic representation of the physical set-up is shown in the Figure~\ref{rbc:setup}. In the rotating frame, we neglect the effect of centrifugal force which is very small compared to the gravity ${\bf{g}}\equiv (0,0,g)$.

\subsection{Governing equations and Parameters }
The convective motion of the fluid in the above described system, under Boussinesq approximation~\citep{boussinesq:1903}, is then governed by the following set of dimensionless equations
\begin{eqnarray}
\frac{\partial {\bf{u}}}{\partial t} + ({\bf{u}}\pmb{\cdot \nabla}){\bf{u}} &=& -\pmb{\nabla}{\pi} + \nabla^2{\bf{u}} + \mathrm{Ra} \theta {\bf{\hat{e}}}_3+ \sqrt{\mathrm{Ta}}({\bf{u}} \times {\bf{\hat{e}}}_3)\nonumber \\ && +\mathrm{Q}\left[\frac{\partial{\bf b}}{\partial z} + \mathrm{Pm}({\bf b}\pmb{\cdot \nabla}){\bf b}\right], \label{eq:momentum} \\   
\mathrm{Pm}[\frac{\partial{\bf b}}{\partial t} + ({\bf u}\pmb{\cdot \nabla}){\bf b} &-& ({\bf b}\pmb{\cdot \nabla}){\bf u}] = {\nabla}^2 {\bf b} + \frac{\partial{\bf u}}{\partial z},\label{eq:induction}\\      
\mathrm{Pr}\left[\frac{\partial \theta}{\partial t}+({\bf{u}}\pmb{\cdot \nabla})\theta\right] &=& \mathrm{u_3}+\nabla^2\theta, \label{eq:heat}\\
\pmb{\nabla \cdot} {\bf{u}}=0, && \pmb{\nabla \cdot} {\bf{b}}=0.\label{eq:div_free}
\end{eqnarray}
In the above system~[Equations~(\ref{eq:momentum}) -~(\ref{eq:div_free})], ${\bf u}(x,y,z,t)=(u_1,u_2,u_3)$ is the convective velocity field, $\theta(x,y,z,t)$ measures the deviation in temperature field from steady conduction profile, $\textbf{b}(x,y,z,t)=(b_1,b_2,b_3)$ is the induced magnetic field, $\pi(x,y,z,t)$ is the modified pressure field, and ${\bf{\hat{e}}}_3$ is the unit vector in vertical direction anti-parallel to the gravity. The above set of equations are made dimensionless by using the units $d^2/\nu$, $d$, ${B_0\nu}/\lambda$ and ${\beta d\nu}/\kappa$ for time, length, induced magnetic field, and temperature respectively. The non-dimensionalization procedure gives rise to five dimensionless numbers namely the Rayleigh number $\mathrm{Ra}={\alpha (T_l-T_u) g d^3}/{\kappa \nu}$, the Taylor number $\mathrm{Ta}={4 \Omega^2 d^4}/{\nu^2}$, the Chandrasekhar number $\mathrm{Q}={{B_0}^2 d^2}/{\nu \lambda \rho_0}$, $\rho_0$ being the reference density, the Prandtl number $\mathrm{Pr}={\nu}/{\kappa}$, and the magnetic Prandtl number $\mathrm{Pm}={\nu}/{\lambda}$. 

In this paper, our prime interest is to investigate the nonlinear aspects of overstable oscillatory convection which appear in rotating magnetoconvection of low-$\mathrm{Pr}$ fluids. For these fluids, the magnetic Prandtl number is exceedingly small and we consider the quasi-static approximation ($\mathrm{Pm} \rightarrow 0$) which has been widely used to study the magnetoconvective and rotating magnetoconvective instabilities~\citep{roberts_book, arnab:2014, banerjee_PRE:2020}. As a result, the induced magnetic field becomes slave to the velocity field and Equations~(\ref{eq:momentum}) and~(\ref{eq:induction}) reduce to
\begin{eqnarray} 
\frac{\partial {\bf{u}}}{\partial t} + ({\bf{u}}\pmb{\cdot \nabla}){\bf{u}} &=& -\pmb{\nabla}{\pi} + \nabla^2{\bf{u}} + \mathrm{Ra} \theta {\bf{\hat{e}}}_3\nonumber \\ && + \sqrt{\mathrm{Ta}}({\bf{u}}\times {\bf{\hat{e}}}_3)+\mathrm{Q}\frac{\partial{\bf b}}{\partial z}, \label{eq:momentum1} \\   {\mathrm{and}~~~}
\nabla^2\bf{b} &=& -\frac{\partial {\bf{u}}}{\partial z}. \label{eq:magnetic1} 
\end{eqnarray}
The dynamics of the system~[Equations~(\ref{eq:heat}) -~(\ref{eq:magnetic1})] is then controlled by the remaining four dimensionless numbers apart from $\mathrm{Pm}$. 

\subsection{Boundary conditions}
We assume the bottom and top plates located at $z = 0$ and $1$ to be stress-free and perfect conductors of heat. This gives the boundary conditions for the velocity and temperature fields 
\begin{equation}
{u_3} = \frac{\partial {u_1}}{\partial z} = \frac{\partial {u_2}}{\partial z} = \theta = 0~\mathrm{at}~ z = 0,~1.\label{bc1}
\end{equation}
To specify the boundary conditions for the magnetic field, we further assume the bounding plates to be electrically insulating. Therefore, currents cannot cross the horizontal boundaries and the normal component of the current density ${\bf j} = ({1}/{\mu_0})(\pmb{\nabla} \times {\bf b})$ ($\mu_0$ being the permeability of free space) vanishes there. However, the induced magnetic field is continuous at the boundaries and the induced magnetic field ${\bf b}^{ext}$ in an electrically non-conducting plate of permeability $\mu_{ext}$ is equivalent to a vacuum and derivable from a potential $\psi$~\citep{chandra:book}. Thus, at the horizontal boundaries we have ${\bf b} = {\bf b}^{ext}$, where ${\bf b}^{ext} = \pmb{\nabla} \psi$ with $\nabla^2 \psi = 0$. In the limit $\mathrm{Pm} \rightarrow 0$, boundary conditions for the magnetic field are derived using the Equation~(\ref{eq:magnetic1}) and are given by
\begin{equation}
{b_1} = {b_2} = \frac{\partial {b_3}}{\partial z} = 0~\mathrm{at}~ z = 0,~1.\label{bc2}
\end{equation}
This choice is also compatible with the divergent nature of induced magnetic field. However, non-zero horizontal velocity components of the fluid give rise to surface currents at the horizontal boundaries which are fixed using the continuity property of the vertical component of the induced magnetic field.


\section{Linear stability analysis and numerical techniques}\label{sec:lsa_nm}
In this section we discuss about the linear stability analysis of the system and numerical techniques used for the investigation.

\subsection{Linear stability analysis}\label{subsec:lsa}
The linear stability analysis of the system is after~\cite{chandra:book}. By performing normal mode analysis, he determined the expressions of the Rayleigh number and angular frequency ($\mathrm{\sigma}$) as a function of wave number ($k$) for the onset of overstable oscillatory convection, given by
\begin{equation}
\mathrm{Ra}(\mathrm{Ta},\mathrm{Q},\mathrm{Pr})=(\frac{\pi^2+k^2}{k^2}) \Big[(\pi^2+k^2)^2+\mathrm{Q}\pi^2-\mathrm{Pr}\sigma^2+\frac{\mathrm{Ta}\pi^2(\pi^2+k^2)((\pi^2+k^2)^2+\mathrm{Q}\pi^2+\mathrm{Pr}\sigma^2)}{\{(\pi^2+k^2)^2+\mathrm{Q}\pi^2\}^2+\sigma^2(\pi^2+k^2)^2}\Big],	\label{eq:marginalstate}
\end{equation}
\begin{equation}
\mathrm{\sigma}(\mathrm{Ta},\mathrm{Q},\mathrm{Pr})= \bigg[\left( \frac{\mathrm{Ta}\pi^2}{\pi^2+k^2}\right)\left\{\frac{(1-\mathrm{Pr})(\pi^2+k^2)^2-\mathrm{Q}\pi^2\mathrm{Pr}}{(1+\mathrm{Pr})(\pi^2+k^2)^2+\mathrm{Q}\pi^2\mathrm{Pr}}\right\} -\left(\pi^2+k^2+\frac{\mathrm{Q}\pi^2}{\pi^2+k^2}\right)^2 \bigg]^{1/2}.\label{eq:omega}
\end{equation}

\begin{figure}
\begin{center}
\includegraphics[height=8cm, width = 0.8\textwidth]{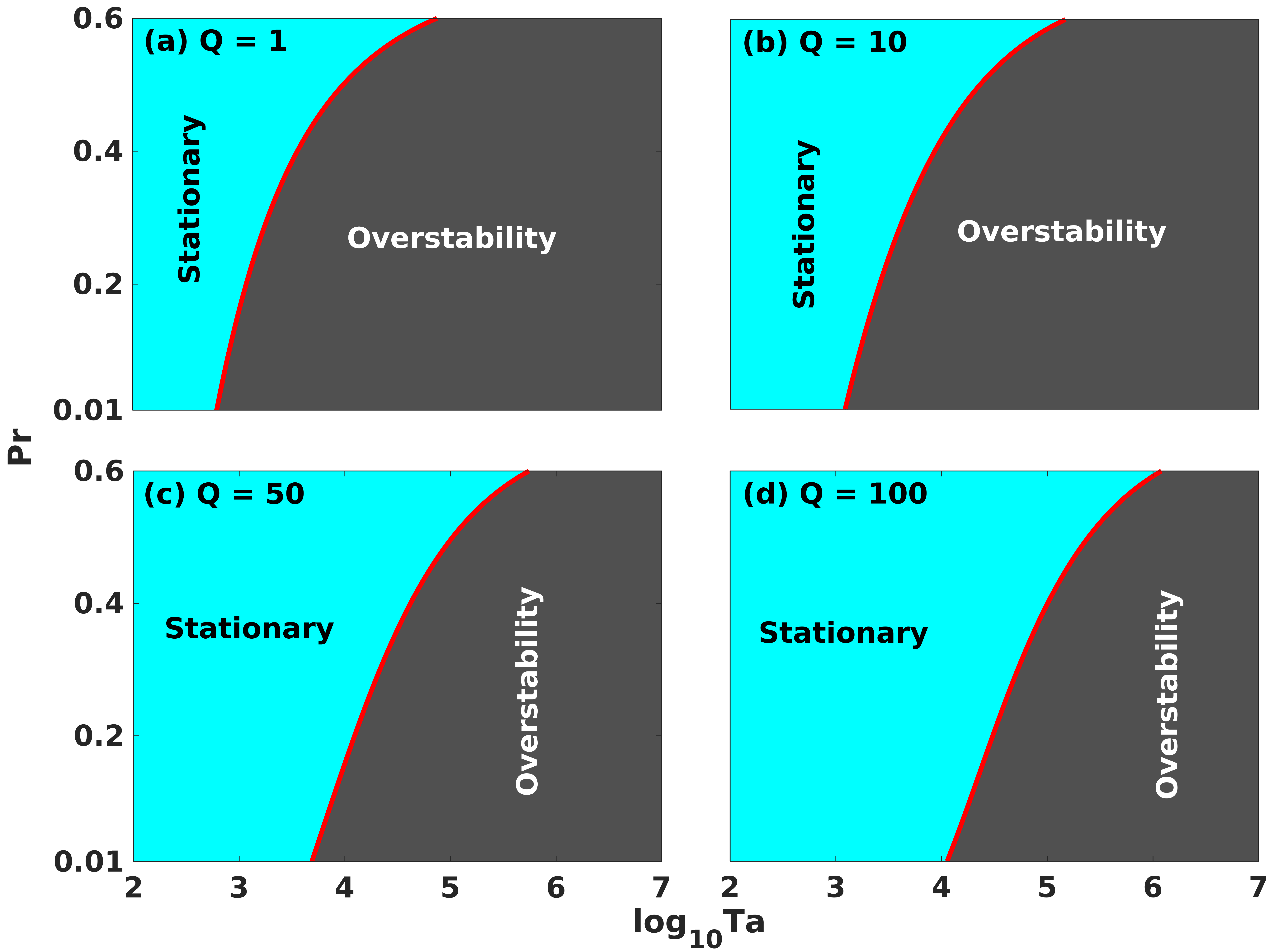}
\caption{Two-parameter $\mathrm{Ta}-\mathrm{Pr}$ plane constructed from linear theory depicting the nature of the onset for four different $\mathrm{Q}$. Cyan and dark gray regions, respectively, represent the parameter space where stationary convection and overstability are preferred at the onset.}
\label{fig:LT_Ta_Pr}
\end{center}
\end{figure}

\begin{figure}
\begin{center}
\includegraphics[height=!, width = 0.8\textwidth]{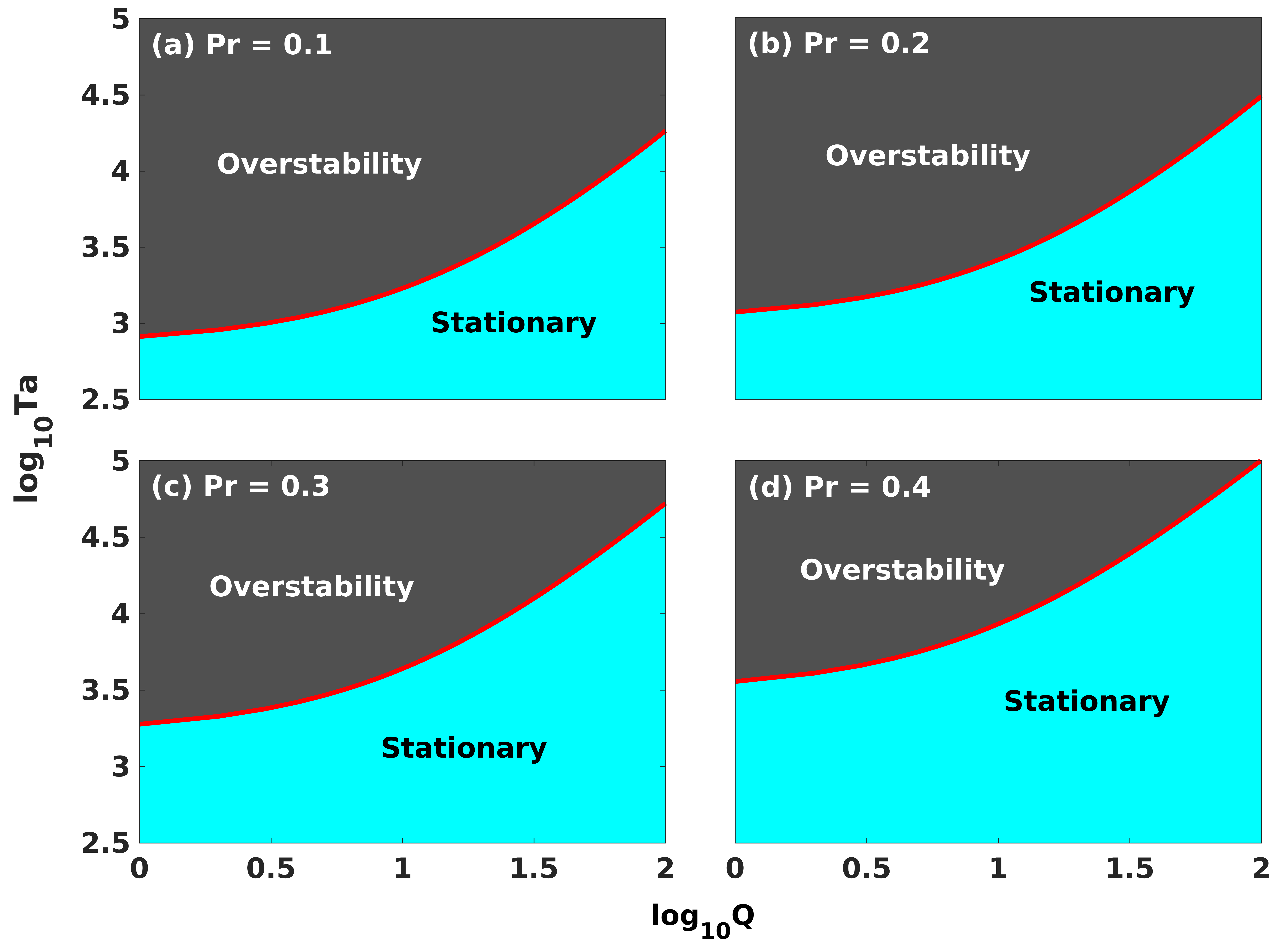}
\caption{Two-parameter $\mathrm{Q}-\mathrm{Ta}$ plane depicting the nature of the onset from the linear theory corresponding to four different $\mathrm{Pr}$. The cyan and dark gray regions as usual represent the parameter space where stationary convection and overstability are preferred at the onset.}
\label{fig:LT_Q_Ta}
\end{center}
\end{figure}

The critical Rayleigh number ($\mathrm{Ra_o}$), associated wave number ($k_o$) and angular frequency ($\sigma_o$) are then determined numerically using the Equations~(\ref{eq:marginalstate}) and~(\ref{eq:omega}) for given $\mathrm{Ta}$, $\mathrm{Q}$ and $\mathrm{Pr}$. We notice that unlike RMC with horizontal magnetic field~\citep{banerjee_PRE:2020}, all three $\mathrm{Ra_o}$, $k_o$ and $\sigma_o$ strongly depend on $\mathrm{Q}$ in this case. We further construct two-parameter $\mathrm{Ta} - \mathrm{Pr}$ plane for different $\mathrm{Q}$ (Figure~\ref{fig:LT_Ta_Pr}) and $\mathrm{Q} - \mathrm{Ta}$ plane corresponding to different $\mathrm{Pr}$ (Figure~\ref{fig:LT_Q_Ta}) to identify the parameter space where overstable oscillatory convection is preferred at the onset. It is clear from Figure~\ref{fig:LT_Ta_Pr} that as $\mathrm{Q}$ increases, the boundary separating the stationary and overstable oscillatory convection regions moves towards the higher $\mathrm{Ta}$ demonstrating the positive role of external magnetic field in suppressing the overstable oscillatory onset. On the other hand, Figure~\ref{fig:LT_Q_Ta} illustrates that the overstable oscillatory onset is also postponed to the higher $\mathrm{Ta}$ as we increase $\mathrm{Pr}$. 

\subsection{Direct numerical simulation}\label{subsec:DNS}
We perform direct numerical simulation (DNS) of the governing Equations~(\ref{eq:heat}) -~(\ref{eq:magnetic1}) together with the boundary conditions~(\ref{bc1}) and~(\ref{bc2}) using an open-source code Tarang based on pseudo-spectral method~\citep{mkv:code}. In the simulation code, vertical velocity, vertical vorticity ($\omega_3$) and temperature are expanded using the set of orthogonal basis functions compatible with the boundary conditions as
\begin{eqnarray}
{u_3}(x,y,z,t)&=&\sum_{l,m,n} W_{lmn}(t)e^{i(lk_ox+mk_oy)}\sin{(n\pi z)},\nonumber\\
\omega_3 (x,y,z,t) &=& \sum_{l,m,n} Z_{lmn}(t)e^{i(lk_ox+mk_oy)}\cos{(n\pi z)},\nonumber\\
\theta (x,y,z,t) &=& \sum_{l,m,n} \Theta_{lmn}(t)e^{i(lk_ox+mk_oy)}\sin{(n\pi z)}.
\end{eqnarray}
The coefficients $W_{lmn}$, $Z_{lmn}$ and $\Theta_{lmn}$ are the Fourier modes with $l$, $m$ and $n$ being the non-negative integers. Components of the velocity and vorticity in the transverse plane are then determined using the solenoidal property of the velocity and vorticity fields. The induced magnetic field is determined from the Equation~\ref{eq:magnetic1}. The scalar potentials at the horizontal boundaries outside fluid~\citep{arnab:2014, ghosh_IJHMT:2020} then can be expressed as
\begin{eqnarray}
\psi|_{z \geq 1} &=& \sum_{l,m,n} (-1)^{(n+1)} \psi_{lmn}(t)e^{ik_o(lx+my)}e^{\gamma(1-z)},\\
\psi|_{z \leq 0} &=& \sum_{l,m,n} \psi_{lmn}(t)e^{ik_o(lx+my)}e^{\gamma z},
\end{eqnarray}
where $\gamma = k_o\sqrt{l^2 + m^2}$ and $\psi_{lmn}(t) = \frac{n\pi W_{lmn}(t)}{\gamma(\gamma^2 + n^2 \pi^2)}$. Simulations are performed in a square box of size $L_x \times L_y \times 1$, where $L_x=L_y=2\pi/k_o$ using random initial conditions. We use $64^3$ spatial grid resolution. Fourth order Runge-Kutta (RK4) integration scheme with Courant-Friedrichs-Lewy (CFL) condition is used for the time advancement with time step $\delta t = 0.001$.

Before performing extensive DNS in our considered parameter space, we validate the DNS results by comparing it with the existing linear theory (LT) results. We first calculate $\mathrm{Ra_o}$, $k_o$ and $\sigma_o$ from LT corresponding to some typical $\mathrm{Ta}$ and $\mathrm{Q}$ for $\mathrm{Pr} = 0.1$. Then we use the $k_o$ determined from LT and perform DNS for the same set of parameter values to determine $\mathrm{Ra_o}$ and $\sigma_o$. Table~\ref{table:DNS_LT_comparison} shows the comparison of these values obtained from LT and DNS. We find a close agreement between LT and DNS results in our considered parameter space. 
\begin{table}
\centering 
\begin{tabular}{c c c c c c c c c}
 $\mathrm{Q}$ & $\mathrm{Ta}$ & $\mathrm{k_o}$ & $\mathrm{Ra_o}$ & $\mathrm{Ra_o}$ & $\mathrm{Error}(\%)$ & Frequency ($\sigma_o$) & Frequency ($\sigma_o$) &  $\mathrm{{Error}}(\%)$\\
    & &  & (LT) & (DNS) &  & (LT) & (DNS) & \\
\hline
& $1698$ & $2.691$ & $2152.07$ & $2150$ & $0.096$ & $2.3869$ & $2.3923$ & $0.226$ \\
$10$ & $5 \times 10^3$ & $2.785$ & $2253.87$ & $2250$ & $0.171$ & $6.3829$ & $6.41$ & $0.425$\\
& $10^4$ & $2.906$ & $2394.51$ & $2387$ & $0.313$ & $9.5387$ & $9.5238$ & $0.156$ \\
& $1.5 \times 10^4$ & $3.006$ & $2523.03$ & $2515$ & $0.318$ & $11.7777$ & $11.7647$ & $0.110$\\
\hline
& $18379$ & $4.257$ & $7074.35$ & $7075$ & $0.009$ & $1.5165$ & $1.5106$ & $0.389$\\
$100$ & $2.5 \times 10^4$ & $4.315$ & $7204.39$ & $7206$ & $0.022$ & $6.2314$ & $6.25$ & $0.298$\\
& $5 \times 10^4$ & $4.5$ & $7653.01$ & $7655$ & $0.026$ & $13.0945$ & $13.1234$ & $0.221$\\
& $10^5$ & $4.769$ & $8417.45$ & $8418$ & $0.007$ & $20.4603$ & $20.4082$ & $0.255$\\
\hline
& $872572$ & $8.011$ & $51217.82$ & $51146$ & $0.140$ & $23.8371$ & $23.4742$ & $1.522$ \\
$1000$ & $9 \times 10^5$ & $8.029$ & $51401.58$ & $51335$ & $0.130$ & $24.9244$ & $25$ & $0.303$\\
& $9.5 \times 10^5$ & $8.059$ & $51732.87$ & $51655$ & $0.151$ & $26.7829$ & $26.67$ & $0.422$\\
& $10^6$ & $8.089$ & $52059.53$ & $51975$ & $0.162$ & $28.5117$ & $28.4091$ & $0.360$\\
\end{tabular} 
\caption{Comparison of DNS and linear theory (LT) for the onset of overstable oscillatory convection.}
\label{table:DNS_LT_comparison}
\end{table}

\subsection{A Low dimensional model}\label{subsec:ldm}
We construct a low dimensional model following the procedure described by~\cite{banerjee_PRE:2020}. We choose similar set of modes i.e. $W_{101}$ in vertical velocity, $Z_{101}$ and $Z_{200}$ in vertical vorticity, and $\Theta_{101}$ and $\Theta_{002}$ in temperature to derive a small model containing only five coupled nonlinear ordinary differential equations given by
\begin{eqnarray}
\dot{X}&=&-\frac{1}{b}(aX+cY-dZ),\nonumber \\
\dot{Y}&=&cX-\frac{a}{b} Y-\frac{\pi}{2}XU,\nonumber \\
\dot{Z}&=&\frac{1}{\mathrm{Pr}}X-\frac{b}{\mathrm{Pr}} Z+\pi XV,\nonumber \\
\dot{U}&=&-4k_o^2U+\pi XY,\nonumber \\
\dot{V}&=&-\frac{4\pi^2}{\mathrm{Pr}}V-\frac{\pi}{2}XZ.
\label{model}
\end{eqnarray}
In the above system the variables are $X = W_{101}$, $Y = Z_{101}$, $Z = \Theta_{101}$, $U = Z_{200}$, and $V = \Theta_{002}$ while the coefficients are $b = (\pi^2+k_o^2)$, $a = Q\pi^2+b^2$, $c = \pi\sqrt{Ta}$, and $d = Rak_o^2$. One major difference between the current model and the $3$-mode model presented by~\cite{banerjee_PRE:2020} is the appearance of $\mathrm{Q}$ in the model equations. As a result, we are not able to perform the adiabatic elimination process to reduce the model further. It is interesting to note that similar low dimensional models have also been used earlier to explore the nonlinear properties of magnetoconvection~\citep{knobloch_JFM:1981} and transitions to chaos in double-diffusive convection~\citep{knobloch_JFM:1986}. But here, we use the model to explore the nonlinear properties of overstable oscillatory convection in RMC with vertical magnetic field. For this purpose, we perform extensive bifurcation analysis of the model using an open-source software XPPAUT~\citep{ermentrout_XPP:book} in our considered parameter space which are discussed next together with the support from DNS. We also introduce a new parameter the reduced Rayleigh number ($r = \mathrm{Ra}/\mathrm{Ra_o}(\mathrm{Ta}, \mathrm{Q}, \mathrm{Pr})$) in the subsequent discussions i.e. at $r = 1$ the convection sets in.

\section{Results and discussions}\label{sec:results}
\begin{figure}
\begin{center}
\includegraphics[height=!, width = 0.8\textwidth]{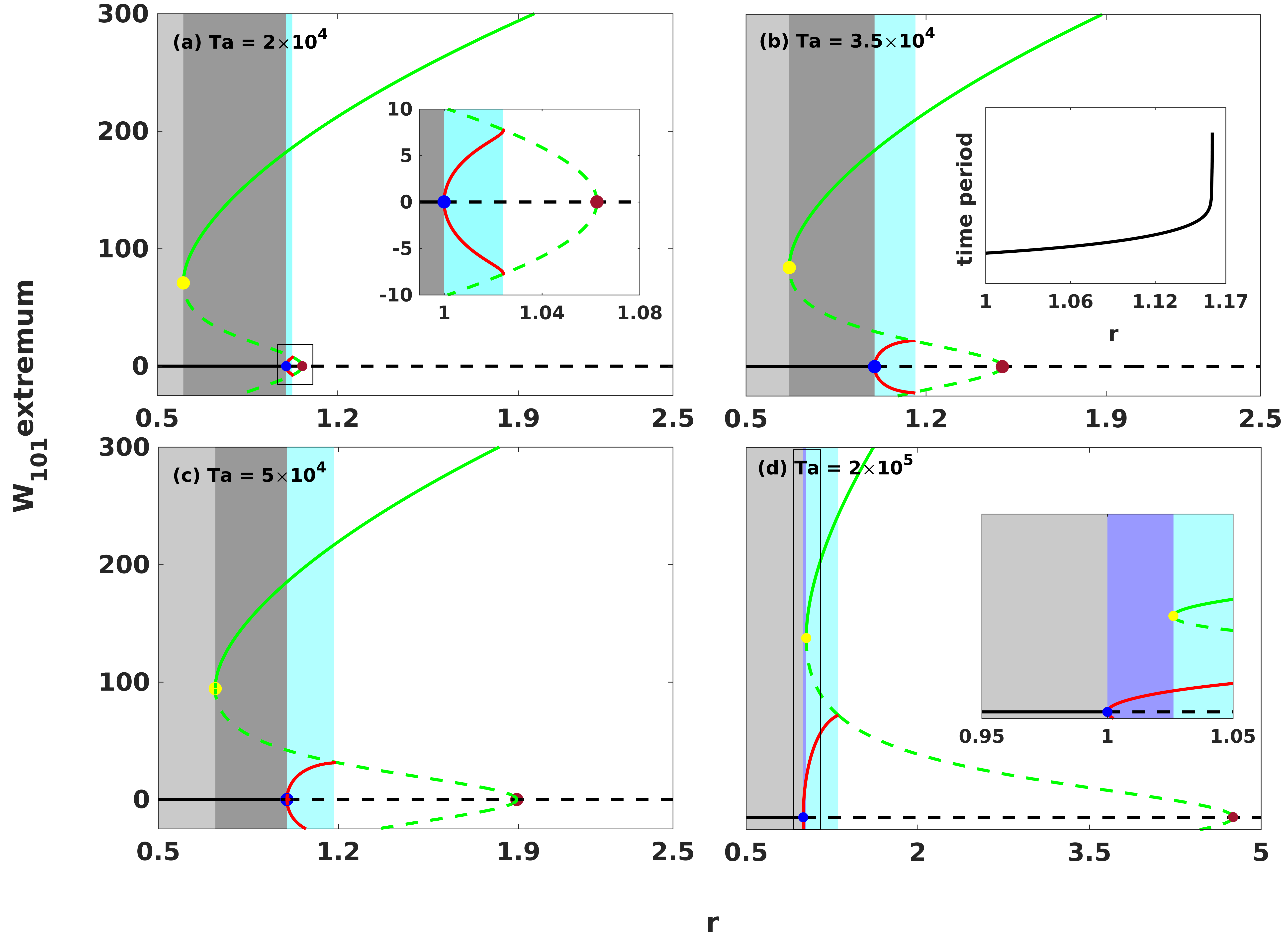}
\caption{Bifurcation diagrams constructed using the model corresponding to four different $\mathrm{Ta}$ for $\mathrm{Q} = 100$ and $\mathrm{Pr} = 0.1$. The solid and dashed curves represent the stable and unstable solutions respectively. Black, green and red curves speak for the trivial conduction state, stationary 2D rolls and limit cycles. Solid blue, brown and yellow circles respectively, denote the Hopf, subcritical pitchfork and saddle-node bifurcation points. Light and dark gray regions are respectively, the conduction regions and the regions of subcritical convection. Cyan regions are the bistable regimes where limit cycles and finite amplitude stationary 2D rolls coexist. The violet region in (d) is the oscillatory regime where limit cycle exists solely. The variation of the time period of limit cycle for $\mathrm{Ta} = 3.5 \times 10^4$ is shown in the inset of (b). Enlarged views of the boxed regions in (a) and (d) are shown at the inset.}
\label{fig:tau_variation}
\end{center}
\end{figure} 

We have already mentioned that four control parameters namely $\mathrm{Ta}$, $\mathrm{Q}$, $\mathrm{Pr}$ and $r$ control the dynamics of the system. Thus, in order to visualize the effect of each parameter we analyze the system by varying any two parameters while keeping the others constant. We first explore the effect of $\mathrm{Ta}$ which is discussed below.

\subsection{Effect of rotation}\label{subsec:Effect_of_rotation}
To unfold the effect of high rotation on overstable oscillatory convection onset, we first prepare four bifurcation diagrams corresponding to four different $\mathrm{Ta}$ for $\mathrm{Q} = 100$ and $\mathrm{Pr} = 0.1$ (Figure~\ref{fig:tau_variation}). In the diagrams, extremum of the dominant Fourier mode $W_{101}$ is shown as a function of $r$ corresponding to the different solutions with different colors.

Figure~\ref{fig:tau_variation}(a) shows the bifurcation diagram for $\mathrm{Ta} = 2\times 10^4$. In the figure, the stable and unstable solutions are shown with solid and dashed curves respectively. The conduction state (black curve) is stable for $r < 1$ and at $r = 1$ it becomes unstable through a supercritical Hopf bifurcation (solid blue circle). As a result, small amplitude stable limit cycle originates there whose extrema are displayed with the red curves. The unstable conduction state continues to exist for higher $r$. At $r = 1.062$, it goes through a subcritical pitchfork bifurcation (solid brown circle) and an unstable stationary 2D rolls branch (dashed green curve) originates there. This 2D rolls branch traverses backward and exists for smaller $r$, even for $r < 1$. At $r = 0.601$, the unstable 2D rolls branch becomes stable via a saddle-node bifurcation (solid yellow circle), changes the direction and exists for higher $r$. Therefore, at the onset we observe stationary 2D rolls with much higher amplitude of subcritical origin which coexists with the small amplitude limit cycle of supercritical origin. With the increment in $r$, the limit cycle grows in size and becomes homoclinic to the unstable 2D rolls branch at $r = 1.024$. The limit cycle cease to exist following this homoclinic bifurcation and only stable 2D rolls is observed for $r > 1.024$. The cyan region ($1 \leq r \leq 1.024$) represents the bistable regime where both oscillatory and stationary solutions coexist depending on the choice of initial conditions.

\begin{figure}
\includegraphics[height=!, width =\textwidth]{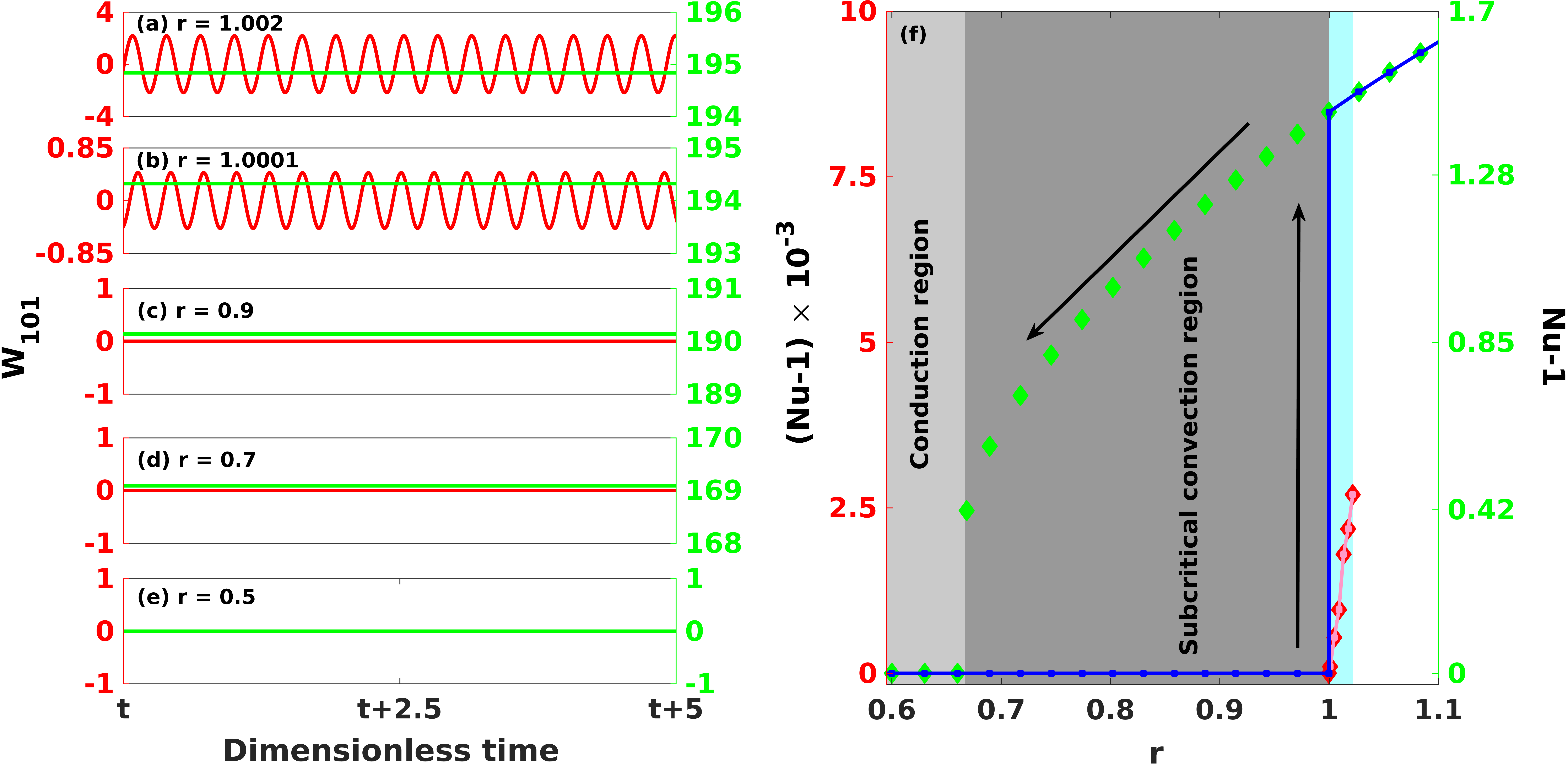}
\caption{Figure shows the simultaneous presence of supercritical convection and subcritical convection accompanied by hysteresis at the onset for $\mathrm{Pr} = 0.1$, $\mathrm{Ta} = 2 \times 10^4$, and $\mathrm{Q} = 100$. (a)-(e) display the time evolution of the largest Fourier mode $W_{101}$ from DNS during the backward numerical continuation. (f) shows the variation of $\mathrm{Nu}$ following the forward and backward numerical continuations corresponding to the small amplitude oscillatory solution and finite amplitude steady solution. Solid blue line with filled squares and green diamonds represent the variation of $\mathrm{Nu}$ during forward and backward continuations of the finite amplitude steady solution while the same for the small amplitude oscillatory solution is shown with solid pink line and red diamonds. Two different vertical scales have been used in each subplot. The vertical scale in the left of each subplot is for the small amplitude oscillatory solutions (red) and that in the right is corresponding to the finite amplitude steady solutions (green).}
\label{fig:heat_transport}
\end{figure}

Due to the presence of stable 2D rolls branch for $r < 1$, we observe appearance of hysteresis there. The convection persists in the dark gray region ($0.6 < r \leq 1$) which is known as the region of subcrtitical convection. The width of this dark gray region which is also the distance between saddle-node and Hopf bifurcation points represents the width of the hysteresis loop. Figure~\ref{fig:heat_transport} depicts the scenario of hysteresis from DNS for $\mathrm{Ta} = 2 \times 10^4$, $\mathrm{Q} = 100$ and $\mathrm{Pr} = 0.1$. We calculate the Nusselt number ($\mathrm{Nu}$, ratio of total heat flux and conductive heat flux) to characterize the hysteresis. Figure~\ref{fig:heat_transport}(f) shows the variation of $\mathrm{Nu}$ calculated following the forward and backward numerical continuations corresponding to the small amplitude oscillatory solution and finite amplitude steady solution. Figures~\ref{fig:heat_transport}(a) -~\ref{fig:heat_transport}(e) display the time evolution of the largest Fourier mode $W_{101}$ for different $r$ during the backward continuation. Note that, random initial conditions are used for the forward continuations while for the backward continuations, the final results of the last simulation are used as the current initial conditions after starting from $r > 1$ with random initial conditions. From Figure~\ref{fig:heat_transport}, we observe that $\mathrm{Nu}$ corresponding to the oscillatory solution shows a continuous change at $r = 1$ during the forward and backward continuations. In contrast, $\mathrm{Nu}$ corresponding to the steady solution exhibits a discontinuous change at the onset during the forward continuation indicating sudden enhancement in heat transport and a hysteresis loop appears during the backward continuation. Therefore, at the onset, we observe simultaneous existence of both subcritical and supercritical convection. It is interesting to note that, the scenario of simultaneous existence of both subcritical and supercritical convection at the onset occurs only when the bistable regime appears there. 

\begin{table}
\begin{center}
\begin{tabularx}{0.9\textwidth}{>{\centering\arraybackslash}X >{\centering\arraybackslash}X >{\centering\arraybackslash}X >{\centering\arraybackslash}X >{\centering\arraybackslash}X}
\multirow{2}{*}{$\mathrm{Ta}~(10^4)$}&\multicolumn{2}{c}{Hysteresis region ($r$)}&\multicolumn{2}{c}{Bistable region ($r$)}\\
 & Model & DNS  & Model  & DNS \\ [1ex]
$2 $ & 0.60 - 1 & 0.67 - 1 & 1 - 1.024 & 1 - 1.022\\
$3.5 $ & 0.67 - 1 & 0.78 - 1 & 1 - 1.151 & 1 - 1.154\\
$5 $ & 0.72 - 1 & 0.88 - 1 & 1 - 1.181 & 1 - 1.235\\
$6.5 $ & 0.76 - 1 & 0.97 - 1 & 1 - 1.205 & 1 - 1.324\\
\end{tabularx}
\end{center}
\caption{Comparison of DNS and model results for $\mathrm{Pr} = 0.1$ and $\mathrm{Q} = 100$.}\label{table:DNS_model_comp_diff_Ta}
\end{table}

\begin{figure}
\includegraphics[height=!, width =\textwidth]{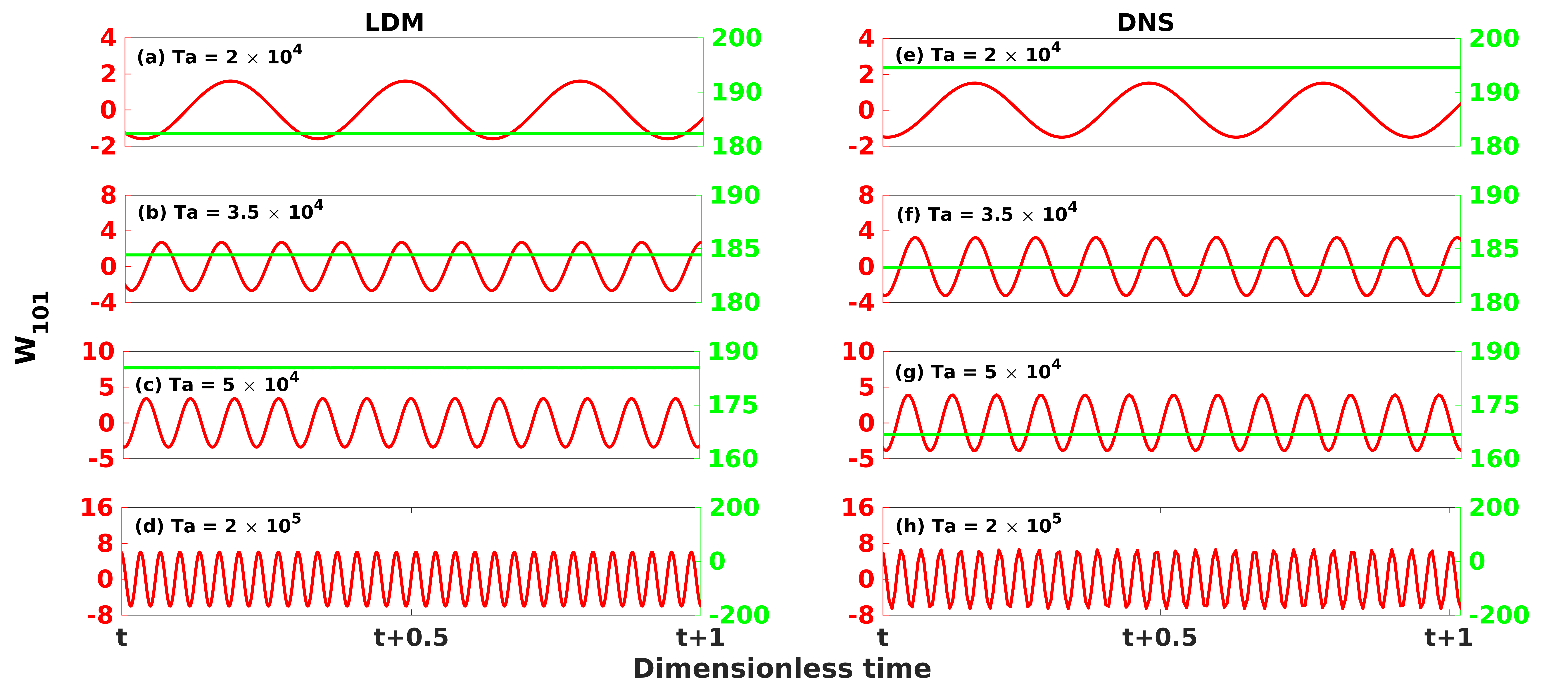}
\caption{Time evolution of the largest Fourier mode $W_{101}$ is shown at the onset of convection ($r = 1.001$) for $\mathrm{Pr} = 0.1$, $\mathrm{Q} = 100$ and four values of $\mathrm{Ta}$ both from the model (left column) and DNS (right column). Two different vertical scales have been used in the figure. The vertical scale in the left of each subplot is for the small amplitude oscillatory solutions (red curves) and that in the right is corresponding to the finite amplitude steady solutions (green curves).}
\label{fig:model_dns_comp_ts}
\end{figure}

Now, to visualize the effect of $\mathrm{Ta}$ we look at the rest of the diagrams (Figures~\ref{fig:tau_variation}(b),~\ref{fig:tau_variation}(c) and~\ref{fig:tau_variation}(d)). From Figures~\ref{fig:tau_variation}(b) and~\ref{fig:tau_variation}(c), we notice that the bifurcation scenario does not differ qualitatively with the increment in $\mathrm{Ta}$, except, the width of the hysteresis loop decreases gradually and that of the bistable regime increases. However, the bifurcation structure differs qualitatively as we increase $\mathrm{Ta}$ further. Figure~\ref{fig:tau_variation}(d) depicts the difference for $\mathrm{Ta} = 2 \times 10^5$. From the figure, we observe that the saddle-node bifurcation point now appears in the convection regime ($r >1$). As a result, the subcritical 2D rolls branch turns ahead of the conduction region and only oscillatory solution prevails at the onset. Due to this, the convection becomes supercritical and the scenario of subcritical convection vanishes. The violet region ($1 \leq r \leq 1.026$) represents the oscillatory regime where only time-dependent solution generated through the Hopf bifurcation exists. The bistable regime shifts towards the higher $r$. DNS provides a good qualitative support in this case with the model results. Table~\ref{table:DNS_model_comp_diff_Ta} displays a qualitative match between the DNS and model results corresponding to different $\mathrm{Ta}$ where width of the hysteresis loop and bistable regime are shown as a function of $r$ for $\mathrm{Q} = 100$ and $\mathrm{Pr} = 0.1$. We further compute the time evolution of dominant Fourier mode $W_{101}$ from the model and DNS near the onset ($r = 1.001$) for $\mathrm{Q} = 100$ and $\mathrm{Pr} = 0.1$ corresponding to four different $\mathrm{Ta}$ (Figure~\ref{fig:model_dns_comp_ts}). We observe a good qualitative agreement between the model and DNS results in this case too. 

\begin{figure}
\includegraphics[height=!, width =\textwidth]{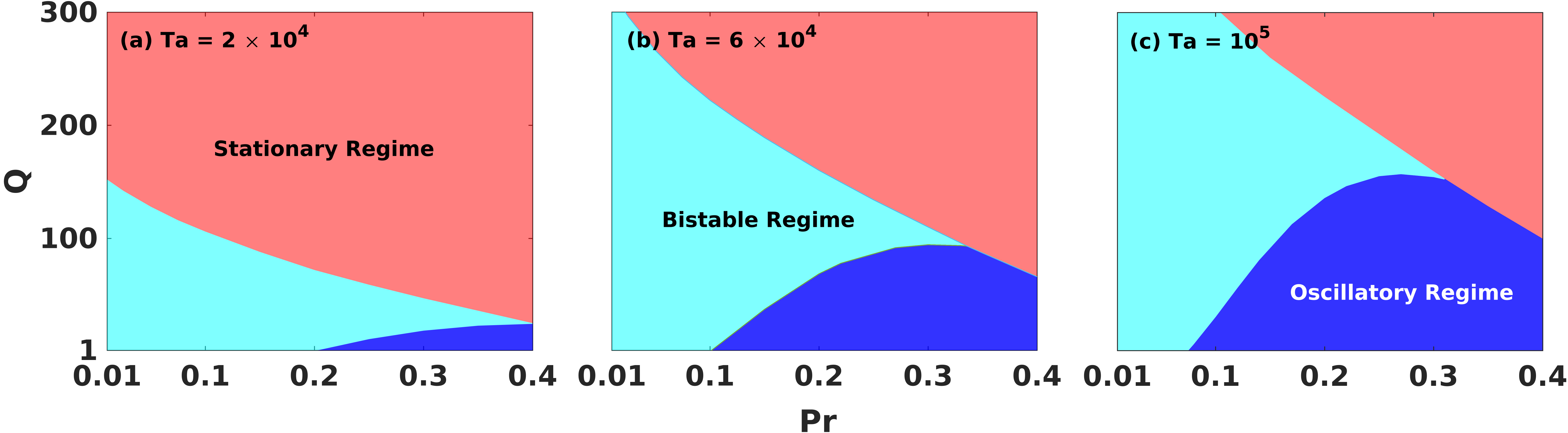}
\caption{Two-parameter $\mathrm{Pr} - \mathrm{Q}$ plane constructed using the model depicting the nature of the onset ($r = 1.001$) corresponding to three different $\mathrm{Ta}$. The orange, cyan and blue regions respectively, represent the stationary regimes, bistable regimes and oscillatory regimes.}
\label{fig:two_par_Pr_Q_diff_Ta}
\end{figure}

We use the model further to identify different regions on two-parameter $\mathrm{Pr} - \mathrm{Q}$ plane for different $\mathrm{Ta}$ depending on the nature of the onset (Figure~\ref{fig:two_par_Pr_Q_diff_Ta}). In the figure, the orange regions represent the parameter space where stationary convection appears at the onset while cyan and blue regions represent the parameter space where bistability and oscillatory solutions are preferred at the onset. It is apparent from the figure that both oscillatory and bistable regimes grow in size with the increment in $\mathrm{Ta}$ while the stationary regime shrinks. Also, the oscillatory solution is preferred at the onset as we increase $\mathrm{Pr}$ while bistability appears for larger $\mathrm{Q}$ with the increment in $\mathrm{Ta}$. Note that, the effect of $\mathrm{Ta}$ here is completely different from that has been reported by~\cite{banerjee_PRE:2020} in RMC with horizontal magnetic field. They found that increment in $\mathrm{Ta}$ favors the subcriticality which solely depends on $\mathrm{Ta}$ and $\mathrm{Pr}$. In contrast, increment in $\mathrm{Ta}$ here kills the subcriticality due to the presence of magnetic field in vertical direction whose effect on the bifurcation structure is discussed next.

\begin{figure}
\begin{center}
\includegraphics[height=!, width = 0.8\textwidth]{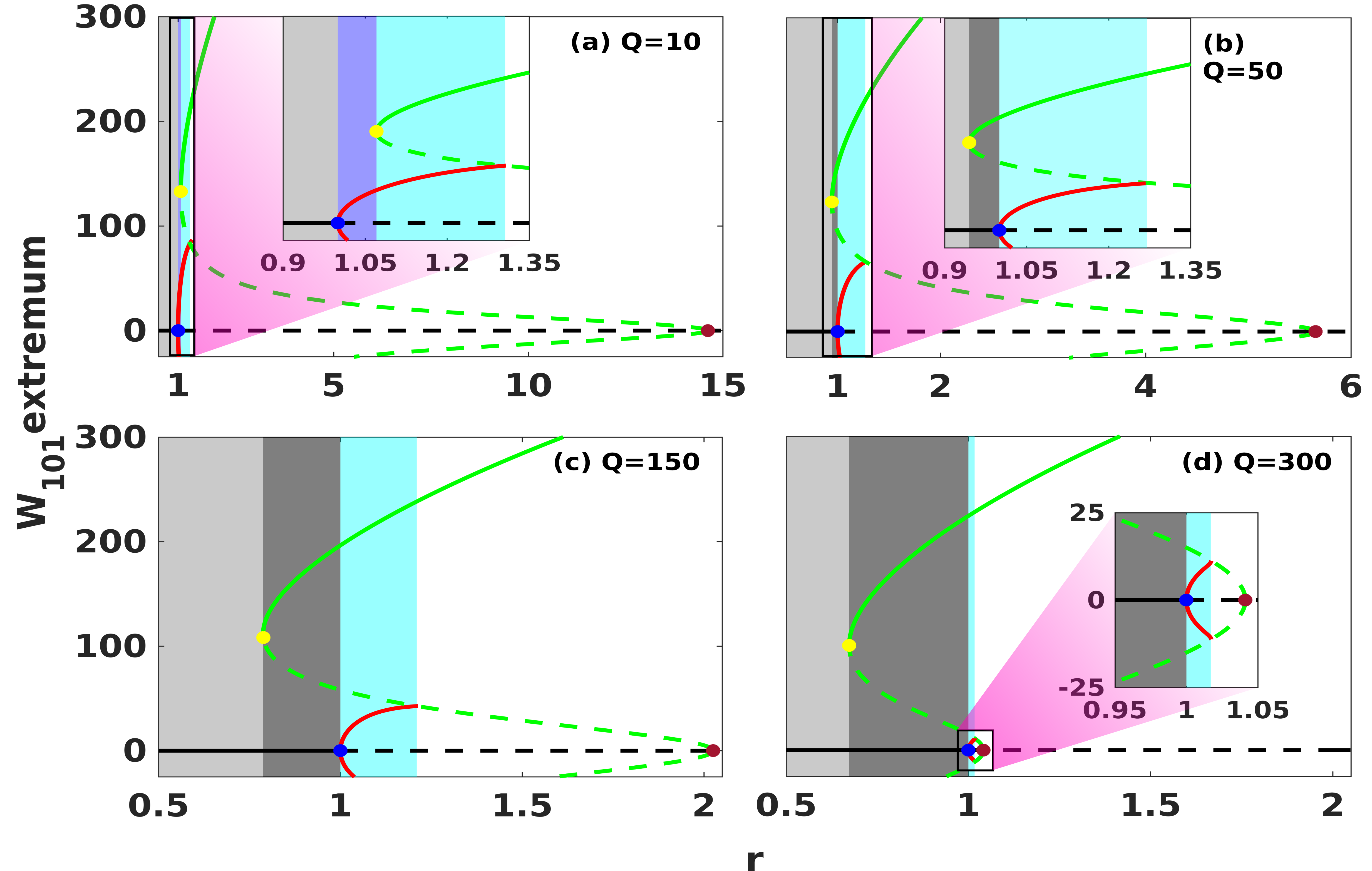}
\caption{Bifurcation diagrams prepared from the model for $\mathrm{Ta} = 10^5$ and $\mathrm{Pr} = 0.1$ corresponding to four different $\mathrm{Q}$. The line style and color coding for different solutions are similar to those used in Figure~\ref{fig:tau_variation}. Zoomed view of the boxed regions are shown in the insets for (a), (b) and (d).}
\label{fig:Q_variation}
\end{center}
\end{figure}

\subsection{Effect of the magnetic field}\label{subsec:effect_of_Q}
We construct four bifurcation diagrams corresponding to four different $\mathrm{Q}$ for $\mathrm{Ta} = 10^5$ and $\mathrm{Pr} = 0.1$ to demonstrate the effect of magnetic field on overstable oscillatory convection (Figure~\ref{fig:Q_variation}). We use similar color coding and line style for different solutions that we have used in Figure~\ref{fig:tau_variation}. From Figure~\ref{fig:Q_variation}, we notice that for relatively weaker magnetic field only oscillatory solution (stable limit cycle) prevails at the onset due to the supercritical Hopf bifurcation (Figure~\ref{fig:Q_variation}(a)). As a result, the convection is supercritical and the bistable regime exists corresponding to the higher $r$. With the increment in $r$, the limit cycle as usual becomes homoclinic to the unstable 2D rolls branch and destroys there. Only stable 2D rolls is observed in the rest of the parameter regime ($r > 1.29$). Qualitative changes appear in the bifurcation structure as we increase the magnetic field's strength. The oscillatory regime gradually shrinks with the increment in $\mathrm{Q}$ and vanishes as $\mathrm{Q}$ crosses a critical value depending on $\mathrm{Ta}$ and $\mathrm{Pr}$. For $\mathrm{Q} = 50$, we observe bistability at the onset i.e. supercritical convection and subcritical convection accompanied by hysteresis both coexist there (Figure~\ref{fig:Q_variation}(b)). The dark gray region as usual represents the hysteresis width where the subcritical convection takes place. With further increment in $\mathrm{Q}$, the width of the bistable regime shrinks and that of the hysteresis grows (Figures~\ref{fig:Q_variation}(c) and~\ref{fig:Q_variation}(d)). Finally, the scenario of overstable oscillatory convection disappears from the system as $\mathrm{Q}$ crosses a critical value $\mathrm{Q} = \mathrm{Q_c}(\mathrm{Ta, Pr}) \approx 303$ predicted by LT. 

\begin{figure}
\includegraphics[height=!, width =\textwidth]{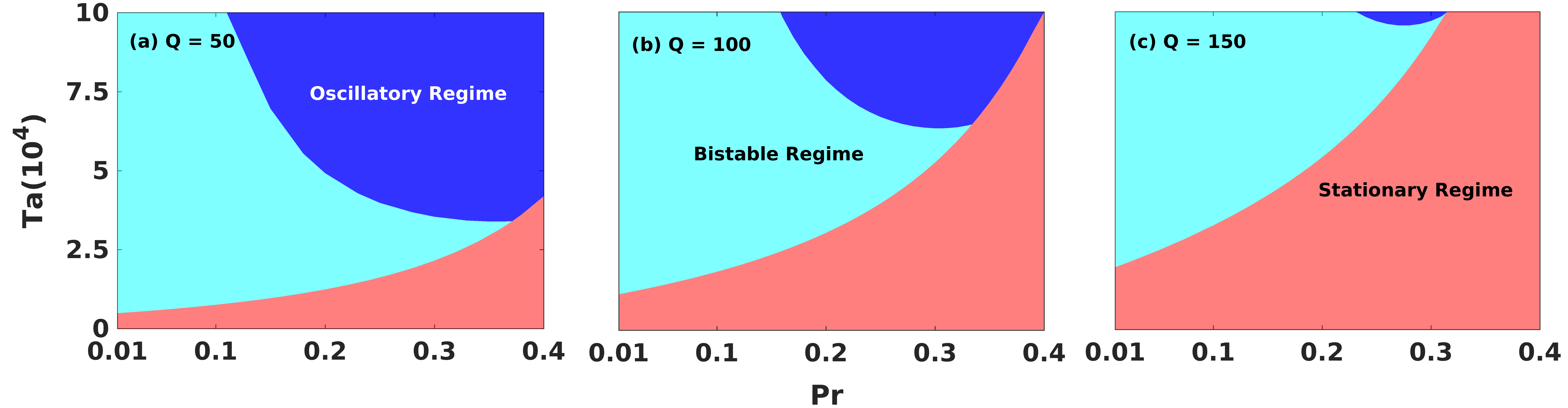}
\caption{Two parameter $\mathrm{Pr} - \mathrm{Ta}$ plane constructed using the model portraying regimes of stationary convection (orange), bistability (cyan) and oscillatory solution (blue) appearing at the onset ($r = 1.001$) corresponding to three different $\mathrm{Q}$.} 
\label{fig:two_par_Pr_Ta_diff_Q}
\end{figure}

The comparison of the model results with the DNS is shown in the Table~\ref{table:DNS_model_comp_diff_Q}. In the table, width of the hysteresis, oscillatory and bistable regimes are obtained as a function of $r$ for different $\mathrm{Q}$ corresponding to $\mathrm{Ta} = 10^5$ and $\mathrm{Pr} = 0.1$ both from the model and DNS. We observe a good qualitative agreement between the DNS and model results. We also portray appearance of different regimes at the onset on two-parameter $\mathrm{Pr} - \mathrm{Ta}$ plane using the model corresponding to different $\mathrm{Q}$ (Figure~\ref{fig:two_par_Pr_Ta_diff_Q}). It is evident from the figure that the stationary regime expands with the increment in $\mathrm{Q}$ while the oscillatory regime shrinks. As a result, occurrence of the oscillatory solution at the onset is postponed to the higher $\mathrm{Pr}$ for fixed $\mathrm{Ta}$ as we increase $\mathrm{Q}$. The bistable regime also moves towards the higher $\mathrm{Ta}$ with the increment in $\mathrm{Q}$. Note that, the effect of $\mathrm{Q}$ here is contradictory from that has been reported by~\cite{ghosh_IJHMT:2020} in the case of stationary convection. They reported that for fixed $\mathrm{Ta}$ and $\mathrm{Pr}$ increment in $\mathrm{Q}$ suppresses the scenario of subcritical convection at the onset and only supercritical convection prevails there. On the contrary, for overstable oscillatory convection, subcriticality is preferred at the onset together with the supercritical convection as we increase $\mathrm{Q}$ for fixed $\mathrm{Ta}$ and $\mathrm{Pr}$. From the above discussion, we also notice that value of $\mathrm{Pr}$ also plays a significant role in determining the dynamics of the system near the onset which we discuss next.

\begin{table}
\begin{center}
\begin{tabularx}{\textwidth}{>{\centering\arraybackslash}X >{\centering\arraybackslash}X >{\centering\arraybackslash}X >{\centering\arraybackslash}X >{\centering\arraybackslash}X >{\centering\arraybackslash}X >{\centering\arraybackslash}X}
\multirow{2}{*}{$\mathrm{Q}$}&\multicolumn{2}{c}{Hysteresis region ($r$)}&\multicolumn{2}{c}{Oscillatory region ($r$)}&\multicolumn{2}{c}{Bistable region ($r$)}\\
 & Model & DNS & Model & DNS & Model & DNS\\ [1ex]
10 & -- & -- & 1 - 1.07 & 1 - 2.08 & 1.07 - 1.30 & 2.08 - 2.56\\
50 & 0.94 - 1 & -- & -- & 1 - 1.38 & 1.00 - 1.26 & 1.38 - 1.82\\
150 & 0.79 - 1 & 0.99 - 1 & -- & -- & 1.00 - 1.20 & 1.00 - 1.29\\
300 & 0.67 - 1 & 0.74 - 1 & -- & -- & 1.00 - 1.02 & 1.00 - 1.02\\
\end{tabularx}
\end{center}
\caption{Comparison of DNS and model results for $\mathrm{Pr} = 0.1$ and $\mathrm{Ta} = 10^5$.}\label{table:DNS_model_comp_diff_Q}
\end{table}

\subsection{Effect of Prandtl number}\label{subsec:effect_of_Pr}
\begin{figure}
\includegraphics[height=!, width =\textwidth]{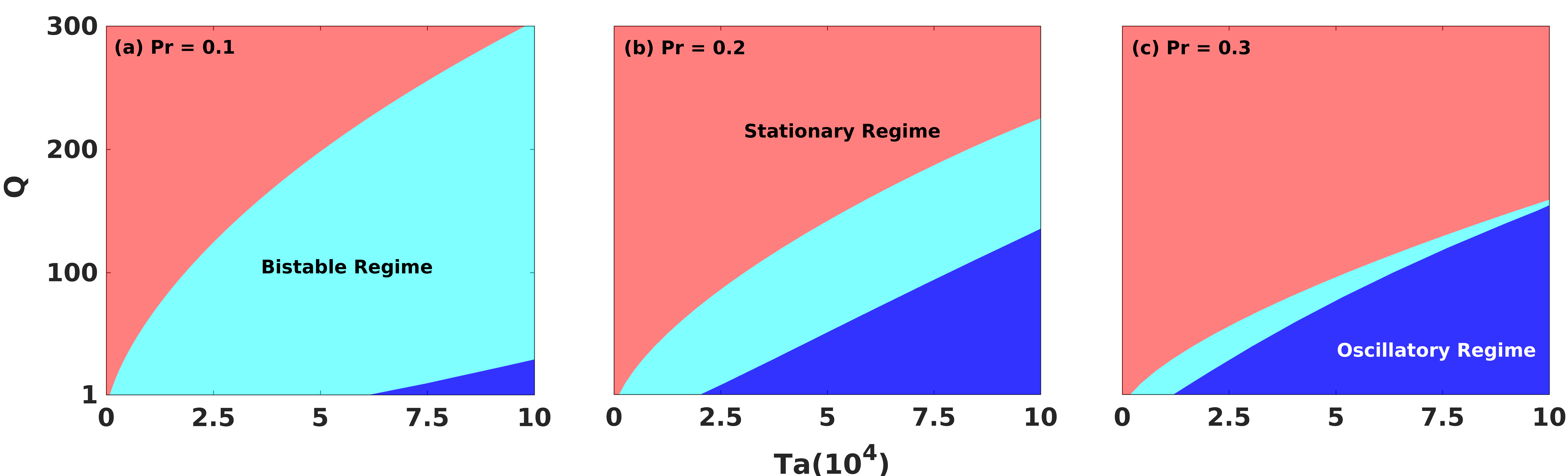}
\caption{Two parameter $\mathrm{Ta} - \mathrm{Q}$ plane constructed using the model corresponding to three different $\mathrm{Pr}$ illustrating the nature of the onset ($r = 1.001$). The color coding is similar that have been used in Figure~\ref{fig:two_par_Pr_Q_diff_Ta} to represent different flow regimes.}
\label{fig:two_para_Ta_Q_diff_Pr}
\end{figure}

To illustrate the effect of $\mathrm{Pr}$ on the bifurcation structure appearing near the onset of overstable oscillatory convection, we first construct two parameter $\mathrm{Ta} - \mathrm{Q}$ diagrams corresponding to three different $\mathrm{Pr}$ (Figure~\ref{fig:two_para_Ta_Q_diff_Pr}). In the figure, emergence of various flow patterns at the onset is portrayed with different colors. From the figure, we observe that both stationary and oscillatory regimes grow in size while the bistable regime shrinks with the increment in $\mathrm{Pr}$. Thus, for fixed $\mathrm{Q}$, with the increment in both $\mathrm{Ta}$ and $\mathrm{Pr}$, the supercritical convection is preferred at the onset and oscillatory solution prevails there. 

\begin{figure}
\begin{center}
\includegraphics[height=!, width = 0.8\textwidth]{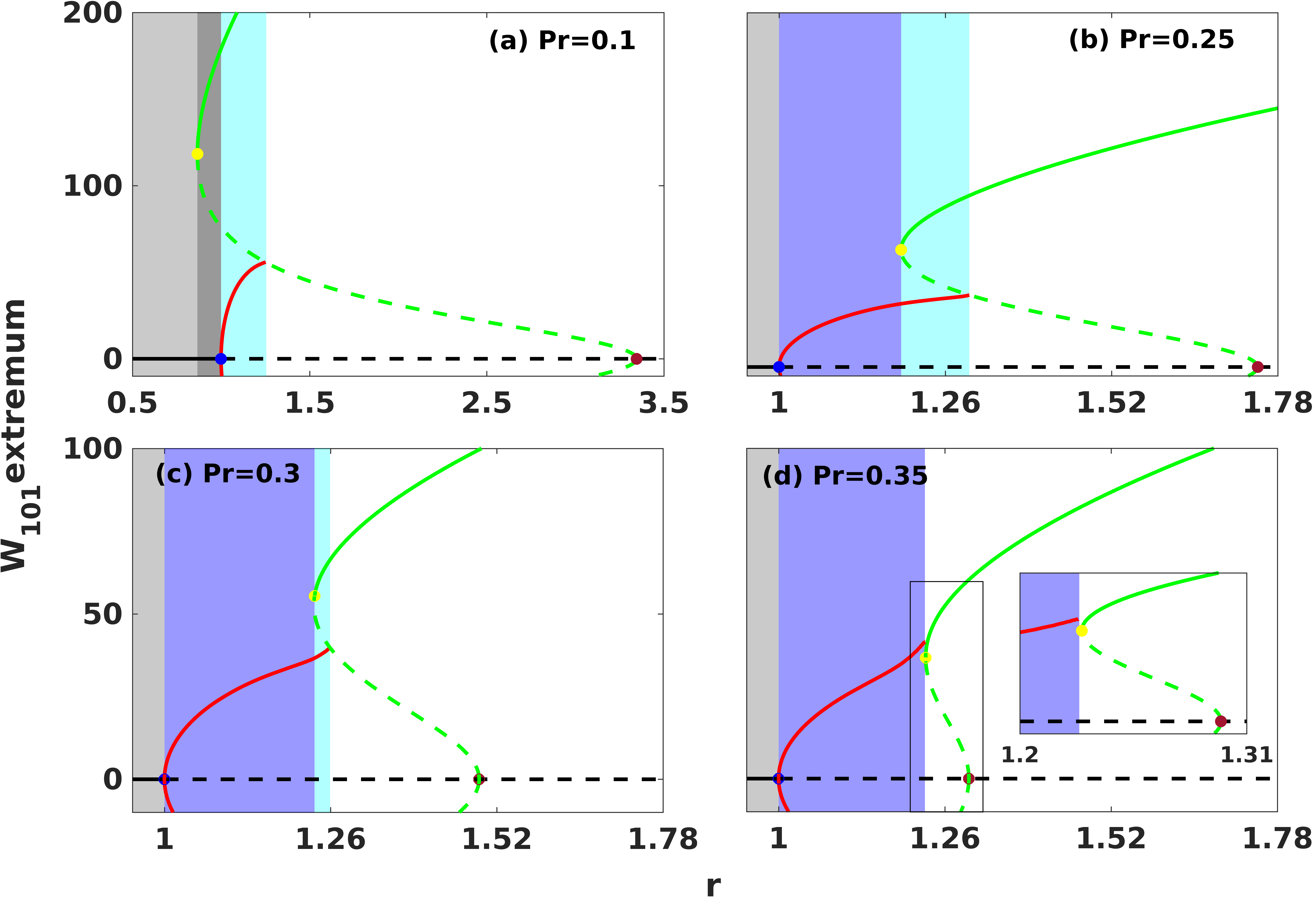}
\caption{Bifurcation diagram constructed using the model corresponding to four different $\mathrm{Pr}$ for $\mathrm{Ta} = 10^5$ and $\mathrm{Q} = 90$. The line style and color coding for different solutions are similar to those used in Figure~\ref{fig:tau_variation}. Enlarged view of the boxed region in is shown in (d) the inset.}
\label{fig:Pr_variation}
\end{center}
\end{figure}

We further construct four bifurcation diagrams using the model by varying $\mathrm{Pr}$ for $\mathrm{Ta} = 10^5$ and $\mathrm{Q} = 90$ to demonstrate the effect of $\mathrm{Pr}$ on the bifurcation structure (Figure~\ref{fig:Pr_variation}). We use similar color coding and line style for different solutions those we have used in Figure~\ref{fig:tau_variation}. From Figure~\ref{fig:Pr_variation}, we notice that for relatively low $\mathrm{Pr}$ bistability is preferred at the onset (Figure~\ref{fig:Pr_variation}(a)). As a result, simultaneous existence of subcritical and supercritical convection is observed there. With the increment in $\mathrm{Pr}$, the scenario of subcritical convection at the onset gets suppressed as the subcritical 2D rolls branch turns well ahead of the conduction region (Figure~\ref{fig:Pr_variation}(b)). Due to this, the bistable regime moves towards the higher $r$ and the oscillatory regime emerges at the onset. With further increment in $\mathrm{Pr}$, the width of the bistable regime decreases and that of the oscillatory regime increases (Figure~\ref{fig:Pr_variation}(c)). Finally, for $\mathrm{Pr} = 0.35$, the bistable regime disappears from the system and only oscillatory regime exists (Figure~\ref{fig:Pr_variation}(d)). The limit cycle in this case terminates following an infinite period bifurcation near $r = 1.228$ and only stable 2D rolls is observed in the rest of the parameter range ($r > 1.228$). The width of the oscillatory regime also decreases as we increase $\mathrm{Pr}$ further. The scenario of overstable oscillatory convection eventually disappears from the system as $\mathrm{Pr}$ crosses a critical value $\mathrm{Pr_c} = 0.667$ depending on both $\mathrm{Ta}$ and $\mathrm{Q}$ predicted by LT. DNS also exhibits similar qualitative results.

\section{Conclusions}\label{sec:conclusion}
To conclude, in this paper, we have presented a systematic study of the plane layer Rayleigh-B\'{e}nard convection in simultaneous presence of rotation about the vertical axis and external uniform vertical magnetic field with stress-free boundaries. This simplified model is believed to serve as a crude idealization of the planetary interiors. We focus on the flow dynamics appearing near the onset of convection where overstable oscillatory motion prevails. Three dimensional (3D) direct numerical simulation (DNS) of the governing equations and low dimensional modeling of the system are performed for this purpose. We choose a wide parameter space for the investigation by varying the control parameters namely, the Taylor number ($\mathrm{Ta}$), the Chandrasekhar number ($\mathrm{Q}$) and the Prandtl number ($\mathrm{Pr}$) in the ranges of $750 \leq \mathrm{Ta} \leq 10^6$, $0 < \mathrm{Q} \leq 1000$ and $0 < \mathrm{Pr} \leq 0.5$. The linear theory in this parameter space states that the overstability occurs as the Rayleigh number ($\mathrm{Ra}$) passes a critical value depending on $\mathrm{Ta}$, $\mathrm{Q}$ and $\mathrm{Pr}$. The ``principle of exchange of stability" becomes invalid there and the convection occurs in the form of purely oscillatory motion. Linear theory also reveals that the scenario of overstable oscillatory convection is postponed to the higher $\mathrm{Ta}$ as both $\mathrm{Q}$ and $\mathrm{Pr}$ are increased. Therefore, for fixed $\mathrm{Ta}$, with the increment in $\mathrm{Q}$ (or $\mathrm{Pr}$), the scenario of overstable oscillatory convection vanishes as $\mathrm{Q}$ (or $\mathrm{Pr}$) crosses a critical value and stationary convection is preferred thereafter. 

In contrast, from DNS, we have identified two qualitative different onsets depending on $\mathrm{Ta}$, $\mathrm{Q}$ and $\mathrm{Pr}$. In the first one, we observe occurrence of bistability at the onset i.e. small amplitude oscillatory solution coexists with the finite amplitude steady solution. As a result, simultaneous existence of supercritical convection and subcritical convection accompanied by the hysteresis is observed there. Appearance of hysteresis also causes a sudden enhancement in heat transport at the onset. On the other hand, for the second one, only small amplitude oscillatory solution is observed at the onset. Thus, the convection is supercritical and we do not observe any hysteresis. 

A convenient low dimensional model containing only five coupled nonlinear ordinary differential equations is derived from the DNS data to get insight into these phenomena in detail. Bifurcation analysis of the model reveals that a supercritical Hopf bifurcation of the conduction state is responsible for the supercritical onset while a subcritical 2D rolls branch originated from a subcritical pitchfork bifurcation of the conduction state is responsible for the subcritical onset. Appearance of finite amplitude steady solution and sudden enhancement in heat transport at the onset are also found to be associated with this subcritical 2D rolls branch. Analysis of the model together with the performance of DNS also ensures that the scenario of subcriticality has strong dependence on all three control parameters $\mathrm{Ta}$, $\mathrm{Q}$ and $\mathrm{Pr}$. However, the effects of $\mathrm{Ta}$ and $\mathrm{Q}$ on the subcritical convection are found to be quite surprising here. Generally, it is found that rotation about the vertical axis favors the subcritical convection~\citep{ghosh_POF:2020, banerjee_PRE:2020, ghosh_IJHMT:2020}, but here we observe completely opposite scenario i.e. increment in $\mathrm{Ta}$ for fixed $\mathrm{Q}$ and $\mathrm{Pr}$ gradually suppresses the subcritical convection. Similarly, increment in $\mathrm{Q}$ for fixed $\mathrm{Ta}$ and $\mathrm{Pr}$ here favors the subcriticality at the onset while it was reported that in the case of stationary convection, increment in $\mathrm{Q}$ eliminates the subcriticality from the system~\citep{ghosh_IJHMT:2020}. Increment in $\mathrm{Pr}$ corresponding to fixed $\mathrm{Ta}$ and $\mathrm{Q}$ eliminates the scenario of subcritical convection from the system too. Finally, we use the model to construct two-parameter diagrams which depict the nature of the onset quite clearly in our considered parameter space.

Thus, we have succeeded in predicting the dynamics appearing near the onset of overstable oscillatory rotating magnetoconvection. However, a considerable amount of work remains to be done; for example, to expand the considered parameter space and refine the model. Investigation is needed to understand the nonlinear saturation of the finite amplitude steady solution and transitions to chaos. Moreover, in this work, we have considered the illustrative stress-free boundary conditions for the velocity field and electrically insulating boundaries for the magnetic field. Study of overstable oscillatory rotating magnetoconvection with other velocity and magnetic boundary conditions is also an open problem and it deserves further investigation. \\

\pagebreak
\noindent {\bf Acknowledgements}\\
P.P. acknowledges support from the Science and Engineering Research  Board  (Department  of  Science  and  Technology, India) (Grant No. MTR/2017/000945). M.G. is supported by the INSPIRE programme of the Department of Science and Technology, India (Code No. IF150261). Authors thank S. Mandal for her useful suggestions and fruitful comments.\\

\noindent {\bf Declaration of interests} \\
\noindent The authors report no conflict of interest.

\bibliographystyle{jfm}
\bibliography{Convection}

\end{document}